\documentclass[sigconf, nonacm]{aamas} 

\usepackage{balance} 
\usepackage{enumitem}
\usepackage{mathtools}
\usepackage{amsthm}
\usepackage{xcolor}
\usepackage[thinc]{esdiff}
\usepackage{amsmath, bm}
\usepackage{dsfont}
\usepackage[font=footnotesize]{caption}
\usepackage[font=footnotesize]{subcaption}
\usepackage{stfloats}

\renewenvironment{proof}{{\textit{Proof sketch.}}}{\qed}

\setcopyright{ifaamas}
\acmConference[AAMAS '25]{Proc.\@ of the 24th International Conference
on Autonomous Agents and Multiagent Systems (AAMAS 2025)}{May 19 -- 23, 2025}
{Detroit, Michigan, USA}{A.~El~Fallah~Seghrouchni, Y.~Vorobeychik, S.~Das, A.~Nowe (eds.)}
\copyrightyear{2025}
\acmYear{2025}
\acmDOI{}
\acmPrice{}
\acmISBN{}

\acmSubmissionID{63}

\title[Learning in all-or-nothing public goods]{Multi-agent reinforcement learning in the all-or-nothing public goods game on networks}

\author{Benedikt Valentin Meylahn}
\affiliation{
  \institution{University of Amsterdam}
  \city{Amsterdam}
  \country{The Netherlands, \today}}
\email{b.v.meylahn@uva.nl}

\begin{abstract}
We study interpersonal trust by means of the all-or-nothing public goods game between agents on a network. The agents are endowed with the simple yet adaptive learning rule, \textit{exponential moving average}, by which they estimate the behavior of their neighbors in the network. Theoretically we show that in the long-time limit this multi-agent reinforcement learning process always eventually results in indefinite contribution to the public good or indefinite defection (no agent contributing to the public good). However, by simulation of the pre-limit behavior, we see that on complex network structures there may be mixed states in which the process seems to stabilize before actual convergence to states in which agent beliefs and actions are all the same. In these metastable states the local network characteristics can determine whether agents have high or low trust in their neighbors. More generally it is found that more dense networks result in lower rates of contribution to the public good. This has implications for how one can spread global contribution toward a public good by enabling smaller local interactions.

\end{abstract}

\keywords{Multi-agent learning, Social dilemma, Trust, Public goods}

\newcommand{\BibTeX}{\rm B\kern-.05em{\sc i\kern-.025em b}\kern-.08em\TeX}

\begin{document}


\pagestyle{fancy}
\fancyhead{}

\settopmatter{printfolios=true}
\maketitle 

\section{Introduction}\label{sec:intro}
\pagenumbering{arabic}
Interpersonal trust is important to the healthy functioning of societies. When trust is in abundance this enables interaction in the absence of costly contracts, monitoring and policing. In this paper we study interpersonal trust as it may be learned by agents in a group setting. To do this we use the public goods game~\cite{Suzuki2005,Santos2008,Wang2009,Szolnoki2010, Tavoni2011, Bladon2011,Szolnoki2011,Dannenberg2015,Javarone2016,Barfuss2020} in the context of agents who learn based on experience. We study the impact of group size and the effect of complex network structure.

The public goods game consists of a group of players who must decide whether or not pay for a public good. Traditionally, the benefit to all players is proportional to the number of players who decide to contribute (or cooperate). Contribution to the public good however, comes at a cost. A selfish agent maximizes their reward by not contributing as long as the others do. In this paper we consider an all-or-nothing version in which the benefit is only rewarded when all players pay the cost.

We are motivated by situations in real life in which each participant has a veto power. Consider for instance the sharing of a water source, if all involved handle it with care and avoid pollution it remains useful. However, if even one firm pollutes the source, this ruins the situation for all others. Indeed, a threshold public goods game is well suited to the modeling of such an ecological conundrum~\cite{Tavoni2011,Dannenberg2015,Barfuss2020}. Another example is slightly reversed though still applicable: Choosing a competitive or a collusive price, can be conceived of as a public goods game. If all players set the collusive price, the `public good' is attained (from the perspective of the firms, for society this is not good at all). However, if but one firm sets the competitive price, the party is ruined for everyone else as they reap the profits of the greatest market share.

In the following sections we provide a non-exhaustive review of the literature on the public goods game. We focus on: all-or-nothing versions, evolutionary approaches, spatial (networked) versions, and experience based agent learning.

\subsection{Literature: General}
Evolutionary dynamics are a prominent tool for scholars studying the evolution of cooperation in social dilemmas. The public goods game is no exception as evidenced by the amount of work in which this is the case (see for example~\cite{Suzuki2005,Santos2008,Wang2009,Szolnoki2010,Szolnoki2011,Javarone2016}). In the evolutionary setting players get rewards for their performance in the game in the form of fitness. Players with a greater fitness are more likely to pass their strategy on to progeny. In these studies long-term stable strategies are identified by means of Monte Carlo simulation or by reducing the system to the replicator equation. 

In the spatial version of the public goods games players are modeled as vertices on a graph and the interactions are defined by connections in the graph. In particular it is common to draw the group of agents for a game by considering the neighborhood of a randomly drawn player. Common network topologies used in the study of the spatial evolutionary public goods game include the torus~\cite{Javarone2016, Wang2024,Lv2024} and grids~\cite{Szolnoki2010,Szolnoki2011}. There has also been work done on evolutionary dynamics of the public goods game on complex networks~\cite{Santos2008,Cao2010,Gomez-Gardenes2011}.

The effect of group size on the possibility of cooperation in the public goods game is an ongoing debate with findings in both empirical and theoretical lines of research going both ways. The interested reader is referred to~\cite{McGuire1974,Isaac1994,Barcelo2015} for a positive influence of group size on cooperation, and \cite{Nosenzo2015, Weimann2022} for negligible and sometimes even negative effect in empirical research). In theoretical research, Suzuki and Akiyama~\cite{Suzuki2005}, for example, show that in the context of evolutionary dynamics well-mixed population including a reputation mechanism, the level of cooperation decreases as the group size increases. In contrast, Szolnoki and Perc~\cite{Szolnoki2011} show that on grids, larger groups lead to more cooperation up to an optimal group size, after which the level of cooperation decreases again. 

\subsection{Literature: All-or-nothing, and learning} \label{sec:aon}
Contribution by all players may be required to avoid disaster, and so we believe the all-or-nothing version of the public goods game to be an important object of study. However, there is not an abundance of literature on this topic. Studying agents that learn and adapt their behavior based on experience is relatively well established for fully-mixed populations (see for example~\cite{Sato2002,Sato2003,Sato2005,Meylahn2024b} or~\cite{Kaniovski1995} for two-player experience based learning). Yet applying experience based learning to agents on a network is still a growing field (see for instance~\cite{Bladon2011,Banisch2019}). As such we now present some the work that relates to these two topics: all-or-nothing public goods, and experience based learning.

Mailath and Postlewaite~\cite{Mailath1990}, and more recently Bierbrauer and Winkelman~\cite{Bierbrauer2020} study the all-or-nothing version of the public goods game from a mechanism design perspective. Mailath and Postlewaite~\cite{Mailath1990} show that for a single public good in which all players have a veto power, 
as the group size grows, the probability of realizing the public good goes to zero. Bierbrauer and Winkelman~\cite{Bierbrauer2020} on the other hand show that if players instead vote on a bundle of public goods, the probability of realizing the public good stays positive dependent on the capacity (size of the bundle). 

Szolnoki and Perc~\cite{Szolnoki2010} consider a version of the game which includes the all-or-nothing version as a special case. In particular they study the effect of a `critical mass' (minimum level of contribution from the group as a whole) required for the benefit to be realised in the spatial evolutionary public goods game on a grid. They find that there is an optimal critical mass yielding a maximum level of cooperation. 

Bladon and Galla~\cite{Bladon2011} study the dynamics of the public goods game with agents (on a ring-and-hub graph) who use batch learning.
They find that increasing the degree of the hub player (connected to all others on the ring) increases their propensity to defect. There are studies that consider experience based learning for agents in the coordination game~\cite{Banisch2019,Meylahn2024b,Meylahn2024c}. In particular the agents in~\cite{Banisch2019,Meylahn2024c} use Q-learning to coordinate opinions with their neighbors on random geometric graphs. Meanwhile in~\cite{Meylahn2024b} agents on a fully connected graph use exponential moving average with a constant learning rate to coordinate in pairwise interactions. In both of these learning methodologies the long-term dynamics exhibit convergence to one strategy (full coordination). 

\subsection{Research gap and contributions}
The advancement of computing technology makes it possible to study a population of sophisticated agents who learn based on past experiences in games. By the nature of social interactions we are motivated to study these on networks of `interesting' topologies. As such in this paper we aim to address this research gap by studying the all-or-nothing public goods game with experience based agent learning for a population on complex networks.

We contribute to multi-agent learning on networks by our theoretical finding that any connected network of agents using the exponential moving average learning rule eventually converges to a single pure strategy. We do this without a separation of time scales which is often used to remove the stochasticity and thereby simplify the analysis (and may result in other dynamics, see~\cite{Borgers1997,Sato2002,Sato2003,Sato2005}). This contributes to understanding dynamics of interacting agents who learn in network environments beyond the calculation of the expected dynamics. 

We also contribute to the understanding of stability of contribution in the public goods game on a network. In particular we show how network properties such as average degree, and network density effect which pure strategy the dynamics converge to. Additionally, we show that on the intermediate time scale, the dynamics may exhibit metastable states in which local network properties strongly influence the behavior of the agents. This highlights the negative effect of interaction size on the likelihood of contributing also in the context of multi-agent learning from experience. We also illustrate that this metastable behavior emerges with network heterogeneity by comparing the dynamics on the random geometric network to those on regular networks (which exhibit fast convergence to steady state).

Our results suggest that to foster contribution on a large scale, it may help to identify local groups where cooperation can yield intermediate benefits. Threading these local groups together without increasing the size of the interaction then might allow this cooperation to spread.


\section{The model} \label{sec:game}
First we introduce the base game as if it were a one shot game. Then we define more formally the repeated and population version.

\subsection{Base game}
The all-or-nothing public goods game consists of $k\in\mathbb{N}$ players acting simultaneously. These players choose whether to contribute ($C$) or defect ($D$). Before making their choice each player $i$ observes a private random variable $\lambda_i\in \mathbb{R}$ which is the payoff they would receive if they and all other players contribute to the public good. For convenience we define the generic random variable such that each $\lambda_i$ is distributed as the generic $\lambda\in\mathbb{R}$. 

The players are not aware of the actions of other players until after all decisions have been made, and they are also not privy to the possible reward obtained by other players ($\lambda_j$ for $j\neq i$). Players that defect (play $D$) obtain a payoff of $1$ regardless of the actions of other players. If all players contribute ($C$), then each player $i$ for $i\in\{1,\ldots,k\}$ obtains the reward $\lambda_i$. Players that contribute while at least one player did not contribute, obtain reward $0$. 

\subsection{Population model of the all-or-nothing public goods game}

We model a population of $N\in \mathbb{N}$ agents who interact in groups $K_t$ of size $k_t\in\mathbb{N}_{\leq N}$ at time steps $t\in\mathbb{N}$. At each time step each agent $i$ in the group $K_t$ observes a random variable\footnote{The idea of each agent observing a randomly drawn utility from the awarded public good is not common yet also not novel. In particular a similar approach is taken in~\cite{Mailath1990,Bierbrauer2020} for the context of mechanism design. In the context of population dynamics and repeated interactions~\cite{Perc2007,Zeng2022,Qian2024} use a random payoff matrix observed by each player in the game drawn anew each round.} (drawn iid) $\lambda_i(t)\in\mathbb{R}$ which determines the size of their gain if the public good is achieved. The agent starts with an endowment of 1 and is required to choose whether or not to contribute 1 to the public good. For each agent $i\in K_t$ we define their action at time $t$:
\begin{equation}
 A_i(t) = \begin{cases}
 C,\quad &\text{if they contribute}\\
 D, &\text{if they defect.}
 \end{cases}
\end{equation}

Subsequently the agent payoff $\pi_c(t)$ defined for taking the contribute action in round $t\in\mathbb{N}$ is\footnote{Awarding the public good only if all agents take the contribute action has been studied for one-shot games in~\cite{Mailath1990,Bierbrauer2020}. The all-or-nothing version we consider is a special case of the threshold version studied experimentally in~\cite{Tavoni2011,Dannenberg2015} and theoretically in~\cite{Wang2009,Szolnoki2010}. These studies however do not consider experience based agent learning or complex network structures as we do.}:
\begin{equation}
\pi_c(t) = \begin{cases}
\lambda_i(t),\quad &\text{if }\sum_{j\in K_t} \mathds{1}_{\{A_j(t)=C\}}=k,\\ 
0,&\text{else}.
\end{cases}
\end{equation}
The payoff for taking the defect action is identically one: $\pi_d(t)=1$.



\subsection{Agent belief}
Each agent $i\in [1,\ldots,N]$ holds a belief on the probability that a random other agent will contribute to the public good $x_i(t)$ (initialized independently for each agent $x_i(0)\sim \mathcal{U}[0,1], \forall i$). For those agents $i\in K_t$ drawn to play the game at time $t$ the belief update follows exponential moving average with fixed learning step size $\alpha$:
\begin{equation}\label{eq:update}
x_i(t+1) = x_i(t)(1-\alpha) + \alpha \left( \frac{\sum_{j\in K_t\setminus i} \mathds{1}_{\{A_j(t)=C\}}}{k_t-1}\right). 
\end{equation}
Agents who were not chosen to interact simply retain their most recent belief. The exponential moving average rule has been studied in the context of two-player games in a series of papers authored by Sato, Akiyama, Farmer, and Crutchfield in various configurations~\cite{Sato2002,Sato2003,Sato2005}. These studies however, make use of a separation of time scales under which the random dynamics converge to the expected dynamics. We study the random dynamics when this separation of time scales does not hold as done for the stochastic coordination game of~\cite{Meylahn2024b}.

\subsection{Agent actions}
The agents don't know the distribution of the reward but only observe the realization of their random variable $\lambda_i(t)$ which they receive if they and all other agents in the round contribute. They combine this with their belief on which actions are likely to be taken by others. This yields the expected utility per action at time $t$ to agent $j$ ($u^c_i(t)$ for cooperating and $u^d_i(t)$ for defecting):
\begin{align}
    u^c_i(t) &=\lambda_i(t) x_i(t)^{k_t-1}\\
    u^d_i(t) & = 1.
\end{align}
We assume that the agents act myopically. That is they chose the action which maximizes their one-round expected utility. Thus, agent $i$ takes the contribute action in round $t$ if $u^c_i(t)\geq u^d_i(t)$:
\begin{equation}
x_i(t)^{k_t-1} \lambda_i(t) \geq  1.
\end{equation}
We use the weak inequality for convenience. Because $\lambda_i(t)$ is continuous, a tie occurs with probability zero, and so this is not an issue.
This can be rearranged to see that the agent contributes if $x_i(t)^{k_t-1} \geq 1/\lambda_i(t)$. For convenience we define the cdf of the random variable $1/\lambda$ derived from the generic random variable $\lambda$ as $F$:
\begin{equation}
F(x):=\mathbb{P}(1/\lambda \leq x),
\end{equation}
and observe that for an outside observer (who does not know the realization of $\lambda_i(t)$ but does know $x_i(t)$) the probability of an agent contributing follows $F(x^{k_t-1})$. Note that $F$ is not associated with any $t$ because the distribution of $\lambda_i(t)$ is always that of $\lambda$ for all $i\in K_t$ and all $t\in \mathbb{R}$. 

An intuitive explanation of the model is provided in Appendix A.

\section{Public goods on networks}\label{sec:nets}
Now we present our results for the stochastic all-or-nothing public goods game with multi-agent learning on networks.

\begin{figure}
    \centering
    \includegraphics[width = 0.35\textwidth]{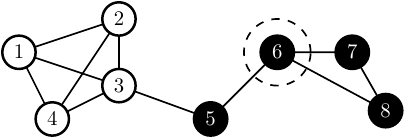}
    \caption{Illustration of how the players are selected to play a round of the all-or-nothing public goods game on a graph of 8 players. The focal player (encircled by a dashed line) is $i=6$. The players taking part in the game are thus $K_t = N[6] = \{6,5,7,8\}$ (indicated by filled nodes) and so $k_t=4$.}
    \label{fig:gameselect}
    \Description{}
\end{figure}

We define a graph $G=(V,E)$ with vertices representing agents $\vert V \vert = N$ and the edges $(u,v)\in E$ representing connections between them. In each discrete round indexed by $t\in\mathbb{N}$, a vertex $i(t)\in V$ (called the `focal' player) is selected uniformly at random from $V$. To be precise $i(t) \sim \mathcal{U}\{1,N\}$ for all $t\in\mathbb{N}$ (there is no correlation between the vertex drawn at round $t$ and $t+1$)\footnote{Introducing a correlation between vertices drawn as the focal vertex, as long as there is positive probability on each vertex to be drawn in each round, does not influence the main result.}. Subsequently the selected player's closed neighborhood $N[i(t)]=\{v: (i(t),v)\in E\}\cup \{i(t)\}$ plays a round of the all-or-nothing public goods game described in \S\ref{sec:game}. This means that the game in round $t\in\mathbb{N}$ with agent $i(t)$ as the focal player has interaction size $k_t=\vert N[i]\vert$ and players $K_t = N[i(t)].$ The way in which agents are selected to take part in a round of the all-or-nothing public goods game on a network is illustrated in Figure~\ref{fig:gameselect}.

\subsection{Main theoretical result}\label{sec:theory}
Now we state the main result which guides our simulations of the model. The result states that as time goes to infinity, the belief and actions of all agents converge to the same action belief pair; either having full trust ($x\to 1$) and always contributing or having no trust ($x\to 0$) and always defecting. This extends a similar result for a fully-mixed (no network) population and pairwise interactions~\cite{Meylahn2024b}.

\begin{proposition}\label{prop:converge}
For learning rate $\alpha\in(0,1)$ and some $\epsilon\in (0,\alpha/d)$ where $d>0$ is the maximum degree in the graph minus one\footnote{Note that for any connected graph with $N>2$ the maximum degree is bigger than 1.}, if  $F(0)=0$, $F(1)=1$, $F(x)\in (0,1)$ for all $x\in (0,1)$ and there exist finite order derivatives of $F(x)$ at $x=0$ and $x=1$ which are finite, then the probability of all agents converging in belief and action asymptotically is one, {\emph i.e.}\ 
\begin{multline}
    \mathbb{P}(\exists t_0: \{\bm{x}(t)\in [0,\epsilon]^N, \forall t\geq t_0\},\\\text{or }\{\bm{x}(t)\in[1-\epsilon,1]^N, \forall t\geq t_0\})=1.
\end{multline} 
\end{proposition}
This proposition means that there can be no disagreement on trust in the very long term. Eventually all agents agree that everyone is to be trusted, or that none is to be trusted.


The proof of this proposition depends on two intermediate results. The first of which states that absorption in the corners is possible:
\begin{lemma}[Absorption in the corners is possible]\label{lem:absorb}
    Let $\alpha\in (0,1)$, $\epsilon\in (0,\alpha/d)$, where $d>0$ is the maximum degree in the graph minus one and suppose $\bm{x}(t_0)\in [0,\epsilon]^N$ for some $t_0\in \mathbb{N}$, then
    \begin{equation}
    \mathbb{P}(\bm{x}(t)\in [0,\epsilon]^N, \forall t\geq t_0 \mid \bm{x}(t_0)\in [0,\epsilon]^N)
    >0.
\end{equation}
For similar $\alpha$ and $\epsilon$, if 
$\bm{x}(t_0)\in [1-\epsilon,1]^N$ for some $t_0\in \mathbb{N}$, then
    \begin{equation}
    \mathbb{P}(\bm{x}(t)\in [1-\epsilon,1]^N, \forall t\geq t_0 \mid \bm{x}(t_0)\in [0,\epsilon]^N)
    >0.
\end{equation}
\end{lemma}
The proof closely follows the proof of Lemma~1 in Meylahn \textit{et al.}~\cite{Meylahn2024b}. We sketch the proof for absorption in $x\in[0,\epsilon]^N$ with attention to the changes due to the network structure and group interaction. 

\begin{proof}
    In order to absorb in the corner $[0,\epsilon]^N$ from round $t_0\in \mathbb{N}$ we require all agents to play defect for all rounds $t\geq t_0.$ We define the probability of the event of all agents playing defect in rounds $t_0, t_0+1,\ldots, t_0+n$ given that all agents start with belief $x\in[0,\epsilon]$ as $D_n$.

    Observe first that any finite connected graph may be covered by cliques which share connections. As such we proceed by induction on the cliques of the network. As base case we consider a graph consisting of one clique of size $k\geq 2$, and call $D_n(k)$ the probability of all $k$ agents playing defect in all rounds $t_0, t_0+1,\ldots, t_0+n$, given that all agents started with belief $x\in[0,\epsilon]$.

    Now for the base case $D_n(k)$ is bounded:
    \begin{equation}
        D_n(k)\geq \prod_{i=0}^n (1-F(z_i^{k-1}))^k,
    \end{equation}
    where $z_0$=$\max_j\{x_j(t_0)\}$, and $z_i = (1-\alpha)z_{i-1}$ for all $i\geq 1$. Subsequently we take the logarithm of both sides and the limit as $n\to\infty$:
    \begin{equation}\label{eq:limDn}
        \lim_{n\to\infty} \log(D_n)\geq \lim_{n\to\infty} k\sum_{i=0}^n\log(1-F(z_i^{k-1})).
    \end{equation}
    We observe that $1-1/y \leq \log(y)$ for all $y\in \mathbb{R}_{>0}$, which we use by setting $1-F(z_i^{k-1}) = y$ to show:
    \begin{equation}
        \log(1-F(z_i^{k-1}))\geq \frac{F(z_i^{k-1})}{F(z_i^{k-1})-1}.
    \end{equation}
     Applying the above to (\ref{eq:limDn}) we get
    \begin{equation}\label{eq:prodseq}
        \lim_{n\to\infty}\log(D_n(k))\geq k\lim_{n\to\infty}\sum_{i=0}^n\frac{F(z_i^{k-1})}{F(z_i^{k-1})-1}.
    \end{equation}
    Proceeding with similar arguments to the proof of Lemma~1 in~\cite{Meylahn2024b},\footnote{This makes use of Abel's convergence test for this product of sequences, and the ratio test with L'H\^{o}pital's rule to show convergence. In this step we require $F$ to have a finite order derivative with a finite value.} we observe that the right hand side of (\ref{eq:prodseq}) converges to a negative number. Taking the exponential of both sides gives the desired result that $\lim_{n\to\infty}D_n(k)>0$.

    As induction step we assume that $L$ cliques of size $k_i\geq 2$ for $i\in\{1,\ldots,L\}$ absorb in the corner at $[0,\epsilon]^N$ with positive probability and call this probability $b_L>0$.

    Once we have shown that $b_{L+1}>0$ then our proof is complete. We construct the induction step graph by referring to its $L$ `original' cliques and the added clique number $L+1$. We remind the reader that the new graph is still connected, thus there are agents in clique $L+1$ who share edges with agents in the $L$ original cliques.
    
    Because we require all agents to play defect in \textit{all} rounds from $t_0$ to $t_0+n$ as $n\to \infty$ we know that each agent will be selected an infinite number of times as the focal agent. Because we are dealing with multiplication we can rearrange the terms from an arrangement by ascending rounds to infinite products of rounds with specific agents as the focal agent. We denote by $k_i$ the number of agents in the game in round $t_0+i$ for $i\in \{0,1,\ldots,n\}$. We define new notation for legibility:
     \begin{equation}
    \mathbb{P}_{[a,b]^k}(\cdot):= \mathbb{P}(\cdot \,| \,x(t_0)\in [a,b]^k).
    \end{equation} 
    Thus by definition of $D_n$ we have:
    \begin{equation}
        D_n = \prod_{i=0}^n\mathbb{P}_{[0,\epsilon]^N}(A(t_0+i)=D\mid A(t)=D,\forall t_0\leq t<t_0+i).
    \end{equation}
    We collect the rounds in which the focal agent is in $a$) the $L$ original cliques and \textit{not} connected also to an agent in clique $L+1$ in set $M_a$, $b$) the clique $L+1$ and \textit{not} connected to an agent in the $L$ original cliques in set $M_b$, and $c$) one of the agents who is connected to both clique $L+1$ and the $L$ original cliques in set $M_c$. So we rewrite the above product:
    \begin{multline}
         D_n = \underbrace{\prod_{t\in M_a}\mathbb{P}_{[0,\epsilon]^N}(A(t)=D\mid A(\tau)=D,\forall t_0\leq \tau<t)}_{=:a} \\
         \times \underbrace{\prod_{t\in M_b}\mathbb{P}_{[0,\epsilon]^N} (A(t)=D\mid A(\tau)=D,\forall t_0\leq \tau<t)}_{=:b} \\
         \times \underbrace{\prod_{t\in M_c}\mathbb{P}_{[0,\epsilon]^N}(A(t)=D\mid A(\tau)=D,\forall t_0\leq \tau<t)}_{=:c}. 
    \end{multline}
    We can split the rounds in $M_c$ further by each agent which, when taking the limit as $n\to \infty$, results in separate infinite products per agent. These are positive by reasoning similar to the base case and noting that each agent's belief is the same or lower than in the base case because they have also played other games (where all agents played $D$ by the conditioning) and so $c>0$. 
    
    In the same way $b>0$ (and $a>0$) because this follows only the agents in clique $L+1$ (the original graph) whose per round belief is thus bounded in the same way as in the base case (or by the induction assumption), with the agents on the boundary who have played in more games which by the conditioning resulted in all agents playing $D$ and thus had even lower beliefs resulting in the defect action at a greater probability.  Thus $a\times b\times c>0.$  
\end{proof}

The other intermediate result states that from a non-corner, the corners may be reached at positive probability. Define $I_N$ as the space of the agent belief excluding the corners, $I_n:=[0,1]^N\setminus \left([0,\epsilon]^N\cup [1-\epsilon,1]^N\right)$.

\begin{lemma}[Reaching the corners is possible]\label{lem:reach}
    Let $\alpha \in (0,1)$ and $\bm{x}(t_0)\in I_N$ for some $t_0\in\mathbb{N}$ and $\epsilon\in (0,\alpha/d)$, where $d>0$ is the maximum degree in the graph minus one, then 
    \begin{equation*}
        \mathbb{P}_{I_N}(\exists t_1\in \mathbb{N}:\bm{x}(t_1)\in [0,\epsilon]^N, t_1>t_0)>0,
    \end{equation*}
    and 
    \begin{equation*}
        \mathbb{P}_{I_N}(\exists t_1\in \mathbb{N}:\bm{x}(t_1)\in [1-\epsilon,1]^N, t_1>t_0)>0,
    \end{equation*}
\end{lemma}
The proof of this lemma closely follows the proof of Lemma~3 in Meylahn \textit{et al.}~\cite{Meylahn2024b}. Here we sketch the differences due to network effects for reaching $\bm{x}\in[0,\epsilon]^N$. The procedure for proving it for $\bm{x}\in[1-\epsilon,1]^N$ is similar.

\begin{proof}
We proceed in two cases. In the first case it follows similarly to the proof of Lemma~3 in~\cite{Meylahn2024b} that the population may reach a corner, while in second case we show that it is possible to reach the starting point of the first case completing the proof.

\textbf{Case 1.} $\bm{x}(t_0)\in [0,1-\epsilon]^N$:
 We first delineate a finite path of positive probability by which a population of $N$ agents reaches the corner given that each agents belief starts in $(0,1-\epsilon)$.
 
    An agent's belief reaches $x\leq \epsilon$, at worst after $\kappa$ rounds with all agents playing $D$ in each round where
    \begin{equation}
        \kappa = \bigg\lceil \frac{\log (\epsilon)}{\log ({1-\alpha})}\bigg\rceil.
    \end{equation}
    For $\alpha$ and $\epsilon$ as defined, this is a finite number. For agents with belief $x\in (0,1-\epsilon)$ the probability of playing defect in a round is positive. 

Consider a vertex cover of the graph and label the agents in it arbitrarily $j=1,\ldots,v$. Suppose we have $t_0$ such that the population belief is $\bm{x}(t_0)\in [0,1-\epsilon]^N$ and in rounds $t_0+i$ for $i=0,1,\ldots, \kappa v$ the focal player $j=i\mod{v}$ is chosen and all agents always play defect in each round. This selection of players happens at probability $(1/N)^{\kappa v}>0$. The probability of all agents always playing defect is also positive. This is because their belief $x(t_0)\leq 1-\epsilon$, and so $(1-F(x_{t_0}^{k_0-1}))^{k_0}>0$, starts positive, and increases each time they have played. They play a finite number of games and thus this finite product of positive numbers converges to a positive number. 

\textbf{Case 2.} $\bm{x}(t_0)\notin [0,1-\epsilon]^N$
Now suppose instead that $\bm{x}(t_0)\notin [0,1-\epsilon]$, implying that at least one agent has belief $x(t_0)>1-\epsilon.$ Suppose as worst case scenario that all agents have belief $x=1$ except agent $j=1$ with belief $x_1=1-\epsilon.$ Now we show that with positive probability this population can reach the state $\bm{x}\in [0,1-\epsilon]$. Let the degree of agent $1$ be $n$ and note that $1-\alpha +\alpha(\frac{n-1}{n})\leq 1-\epsilon$, by the choice of~$\epsilon$.

Agent $1$ is chosen as focal player two rounds in a row at probability $1/N^2$. If in the first of those two rounds, agent $1$ plays defect and all others play contribute, the beliefs of all other agents in the game becomes $x\leq 1-\epsilon$, while agent $1$'s belief becomes $x>1-\epsilon.$ In the second round however the other $n$ agents play defect at positive probability, and so after both rounds, at positive probability $p$, agent $1$ and all of their $n$ neighbors have beliefs $x\leq 1-\epsilon.$ 

Continuing in this way along a spanning tree of the graph, at each iteration playing two games (the first to bring new players into belief $x\leq1-\epsilon$, and the second to ensure that the focal player's belief is also $x\leq1-\epsilon$) each agent in the population has belief $x\leq1-\epsilon$. The largest number of rounds required is for the path graph on $N$ players which results in $2(N-1)$ games. Thus at probability greater than or equal to $1/N^{2(N-1)}p^{N-1}>0$ in at most $2(N-1)$ rounds, all agents have belief $x\leq 1-\epsilon$. From here the proof continues as in case 1.
\end{proof}


Together these results prove Proposition~\ref{prop:converge}. To see this note that after $r<\infty$ rounds, the beliefs of the agents may enter one of the corners at positive probability (by Lemma~\ref{lem:reach}) and be absorbed there (by Lemma~\ref{lem:absorb}). Call the probability of this entering and absorbing event $p_a>0$. Then the probability of never absorbing in one of the two corners behaves like $\lim_{n\to\infty}(1-p_a)^n =0.$

The assumptions on $F$ are minor, and reasonable; if an agent believes there is zero (conversely 100\%) chance of others contributing then they should take the defect (conversely contribute) action with probability one as this surely maximizes their expected reward.

\subsection{Simulation setup}\label{sec:simsetup}
Because Proposition~\ref{prop:converge} only guarantees convergence as $t\to\infty$, by simulation we study the effect of the network on the speed of convergence to steady state and on the pre-limit behavior. We fix the learning rate $\alpha =0.3$ in favor of faster simulation runs. Similarly to balance having interesting network topologies and a manageable computational load we fix the population size $N=50.$ For this numerical simulation we also fix $F(x)=x^{1/4}$.

We set an $\epsilon_s=10^{-4}$, in combination with a maximum number of rounds $T = 10^7$. We stop each iteration if either all agents have belief $x<\epsilon_s$ or $x>1-\epsilon_s$ as proxy for convergence or, alternatively if the maximum number of rounds is reached. 

To study the effect of network topology we use the random geometric network model~\cite{Dall2002,Penrose2003}. This random network model is relevant because it shares properties of actual social networks~\cite{McPherson2001}. The graph is constructed by fixing a radius $r_g\in(0,1)$ and placing $N$ points uniformly at random in the unit square $[0,1]^2$. Subsequently, any two points that are at distance $d< r_g$ are connected by an edge. In particular we focus on connected networks and so for each iteration we draw random geometric graphs repeatedly until we draw a connected geometric graph. For the settings described we run 500 iterations for each setting $r_g \in\{ 0.15,0.2,0.25,0.3\}$.



\subsection{Random geometric network results}\label{sec:results_geo}
\begin{figure}[htb]
    \flushleft
        \includegraphics[width = 0.4\textwidth]{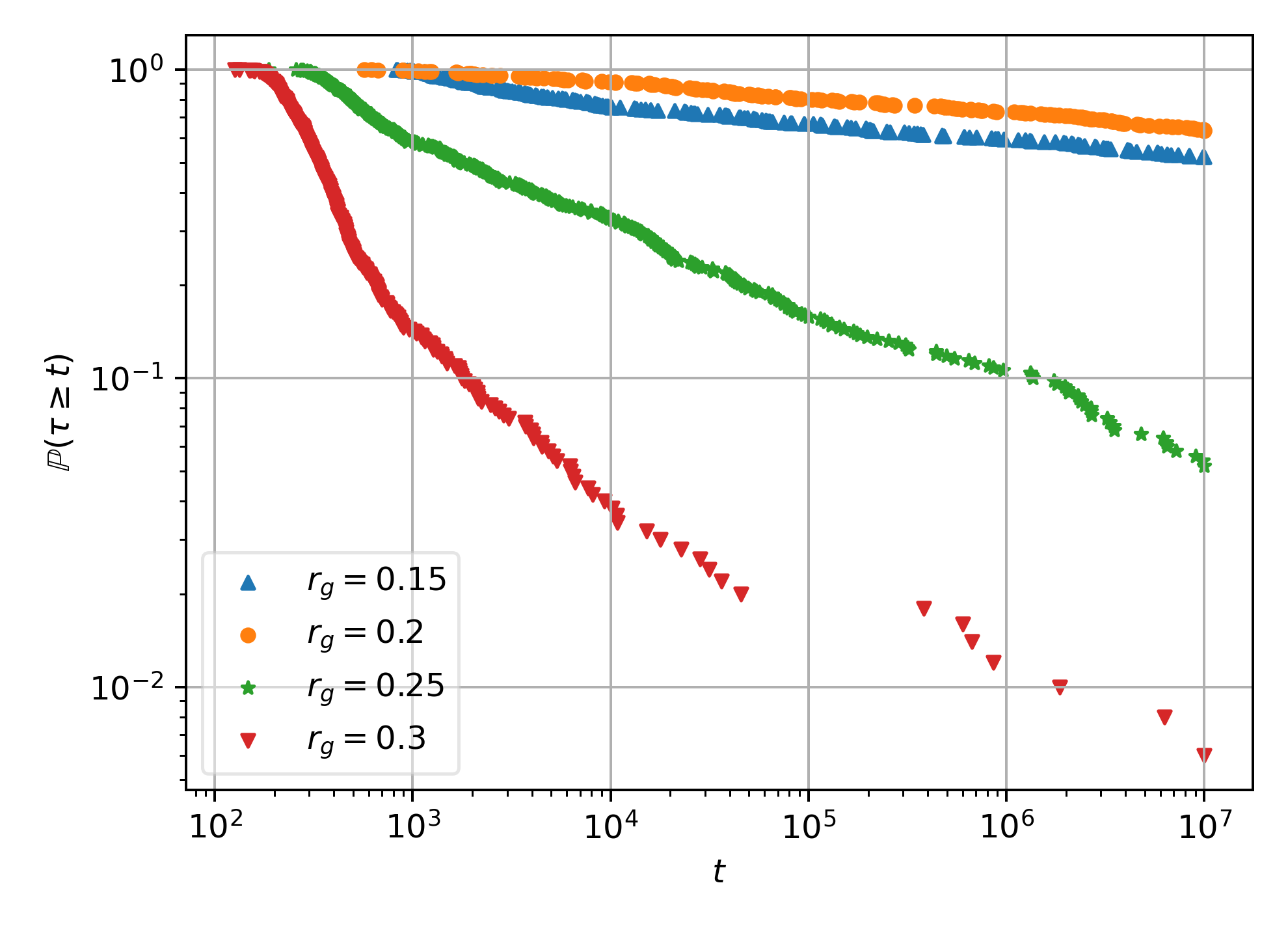}
    \caption{Tail probabilities ($\mathbb{P}(\tau\geq t)$) on a log-log scale for the time to convergence for different values of the radius $r_g$ used in the random geometric graph model to sample networks.}
    \label{fig:con}
\end{figure}

We plot the time to convergence $\tau :=\min\{t:\bm{x}(t)\in [0,\epsilon_s)^N$ or $\bm{x}(t)\in (1-\epsilon_s,1]^N\}$ in Figure~\ref{fig:con} in the form of tail probabilities\footnote{Note that this decision is informed by Proposition~\ref{prop:converge}. If we did not know that convergence happens eventually with probability one, we might falsely plot the probability of convergence rather than the probability of convergence \textit{before} $t\in\mathbb{N}.$} ($\mathbb{P}(\tau\geq t)$). We also tabulate the end state of the simulation runs in Table~\ref{tab:numcon}. In Figure~\ref{fig:con} we notice that increasing the network radius ($r_g$) increases the probability of quicker convergence. Notice that for $r_g=0.15,0.2$ the majority of the runs do not reach consensus in the simulated time which may also be read off Table~\ref{tab:numcon}. 
\begin{table}[htb]
    \centering
    \begin{tabular}{c|c|c|c|c}
         $r_g$ &  $0.15$ & $0.20$ & $0.25$ & $0.3$\\
         \hline
         Contribution & 0.48 & 0 & 0 & 0 \\
         Defection & 0 &  0.366 & 0.948 & 0.994 \\
         Not converge & 0.52 & 0.634 & 0.052 & 0.006
    \end{tabular}
        \caption{Portion of runs out of 500 which converged to contribution and defection as well as the proportion which did not converge in $10^7$ rounds.}
    \label{tab:numcon}
    \vspace*{-2em}
\end{table}
By the linear shape of this tail probability plotted on a log-log scale we posit that a heavy-tailed distribution underlies the time to consensus on a random geometric graph. To check whether this is really the case we turn to the broadest sub-class of heavy-tailed distributions: subexponential distributions (examples are the Pareto, and LogNormal distributions, for an introduction see~\cite{Nair2022}). There is equivalence between distributions satisfying the catastrophe principle and subexponential distributions. A distribution $H$ is said to satisfy the catastrophe principle if for $n\geq 2$
\begin{equation}\label{eq:cat}
    \frac{\mathbb{P}(\max\{X_1,\ldots,X_n\}>t)}{\mathbb{P}(X_1+\cdots+X_n>t)}\to 1,
\end{equation}
as $t\to \infty$ where $X_i$ for $i=1,\ldots,n$ are iid and have distribution $H$. The intuitive understanding of this principle is that a large sum of these random variables is likely caused by a single massive variable\footnote{It has been shown~\cite{Chistyakov1964,Embrechts1997} that if (\ref{eq:cat}) holds for some $n\geq 2$ then it holds for all $n\geq 2$.}. In Table~\ref{tab:cat} we present an estimate of the ratio of (\ref{eq:cat}) for the time to consensus. In particular the ratio is shown for $t=10^6$, and $n=2$. To estimate each ratio we sample pairs of simulations 250 times once with and once without replacement of samples. 
\begin{table}[htb]
    \centering
    \begin{tabular}{c|c|c|c|c}

        $r_g$ &  $0.15$ & $0.20$ & $0.25$ & $0.3$\\         
         \hline
            NR:     & 0.848:0.852 & 0.896:0.896 & 0.196:0.196 & 0.02:0.02 \\
         \hline
            R:     & 0.824:0.824 & 0.9:0.9 & 0.192:0.192 & 0.028:0.028 
    \end{tabular}
    \caption{Ratio between $\mathbb{P}(\max\{X_1,X_2\}>t)$ : $\mathbb{P}(X_1+X_2>t)$ where the $X$s are our simulated time to convergence. NR: no replacement (run $i$ paired with run $i+250$), R: replacement. $t=10^6$, $n=2$.}
    \label{tab:cat}
    \vspace*{-2em}
\end{table}

This is by no means a formal proof of the catastrophe principle holding for our data, though we do see that the ratios in Table~\ref{tab:cat} are 1 or close to 1. The long (and plausibly heavy-tailed) time to convergence would be explained by the existence of metastable state in which the process is `stuck' in for a long time before at some point the process jumps to a true steady state in consensus.

\begin{figure}
    \centering
    \begin{subfigure}{0.25\textwidth}
        \centering \includegraphics[trim={0 5mm 0 3mm}, clip, width = \textwidth]{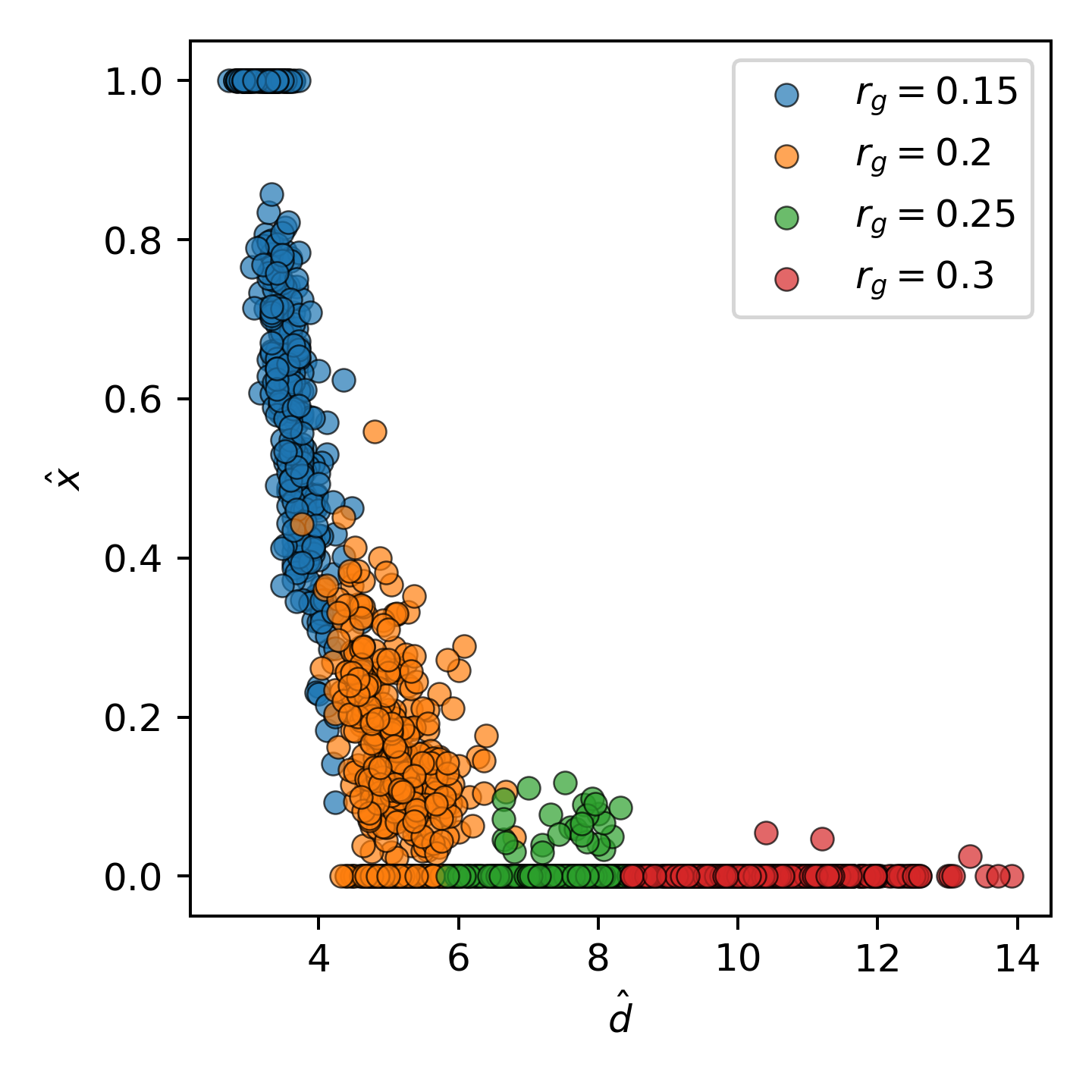}
        \caption{Mean degree}\label{fig:meand}
    \end{subfigure}%
    \begin{subfigure}{0.25\textwidth}
        \centering
         \includegraphics[trim={0 5mm 0 3mm}, clip, width = \textwidth]{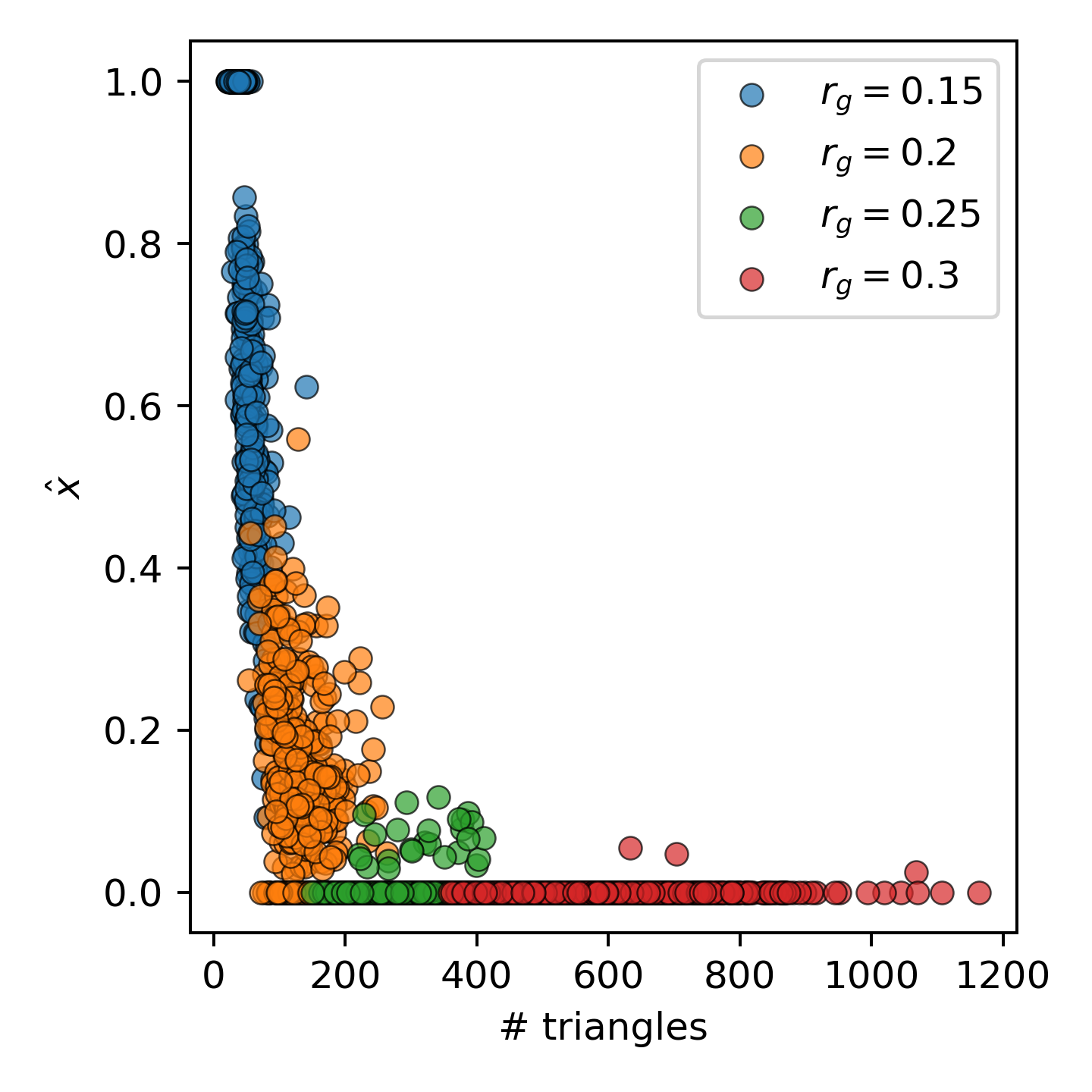}
         \caption{Triangles}\label{fig:tri}
    \end{subfigure}%

    \caption{Average final estimate scatter plotted against various network characteristics. Each point is not fully opaque so that darker regions are indicative of more data points.}
    \label{fig:corr}
\end{figure}

In Figure~\ref{fig:corr} we show scatter plots contrasting network characteristics with the final average agent belief. The color represents the setting of $r_g$. By these plots and Table~\ref{tab:numcon} we see that 
 for $r_g=0.15$ convergence is either to the always-contribute steady state or does not occur. Conversely for all other settings $r_g\in\{0.2, 0.25, 0.3\}$ consensus was reached only on the always-defect steady state or not at all in the simulated time. This seems like a rather abrupt transition based solely on the tabulated results, however, in the scatter plots  we this is somewhat gradual. In general as the $r_g$ increases the final average belief decreases.
 
In Figure~\ref{fig:meand} and~\ref{fig:tri} we see that the mean degree of the network and the number of triangles in the graph have a strong negative correlation to the average agent belief at the end of the simulation run. As the average degree of the network increases, so does the average of the interaction size $k_t$. A similar relationship holds with the number of triangles and the interaction size. It is not surprising that a bigger interaction size $k_t$ should result in more convergence to the always-defect steady state. This is because as $k_t$ increases, $F(x^{k_t-1})$ decreases and so the per turn probability of contributing decreases.



\begin{figure}
    \centering
    \begin{subfigure}{0.25\textwidth}
        \centering
            \includegraphics[trim={0 5mm 0 3mm},clip, width = \textwidth]{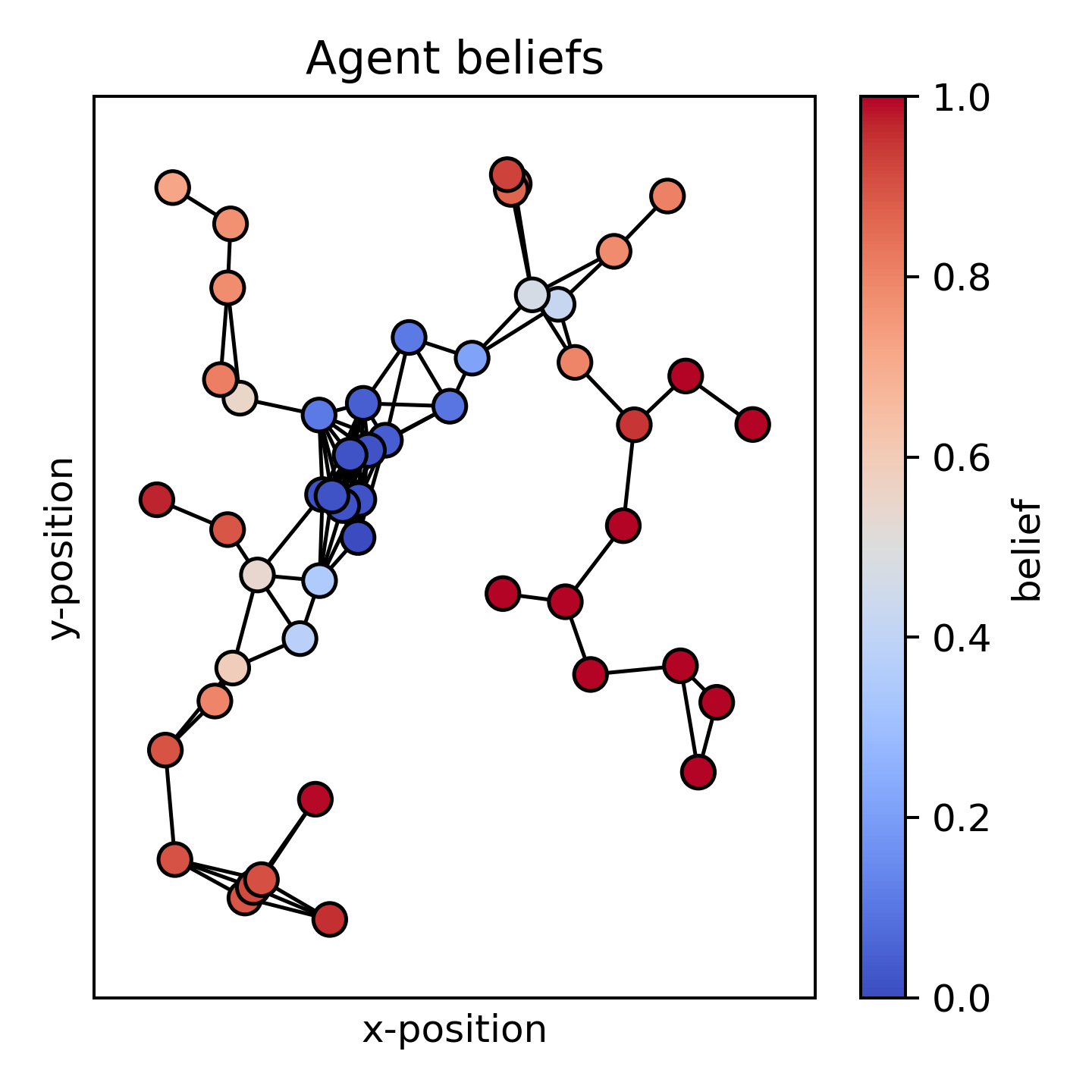}
            \caption{$r_g=0.15$}
    \end{subfigure}%
    \begin{subfigure}{0.25\textwidth}
        \centering \includegraphics[trim={0 5mm 0 3mm}, clip, width = \textwidth]{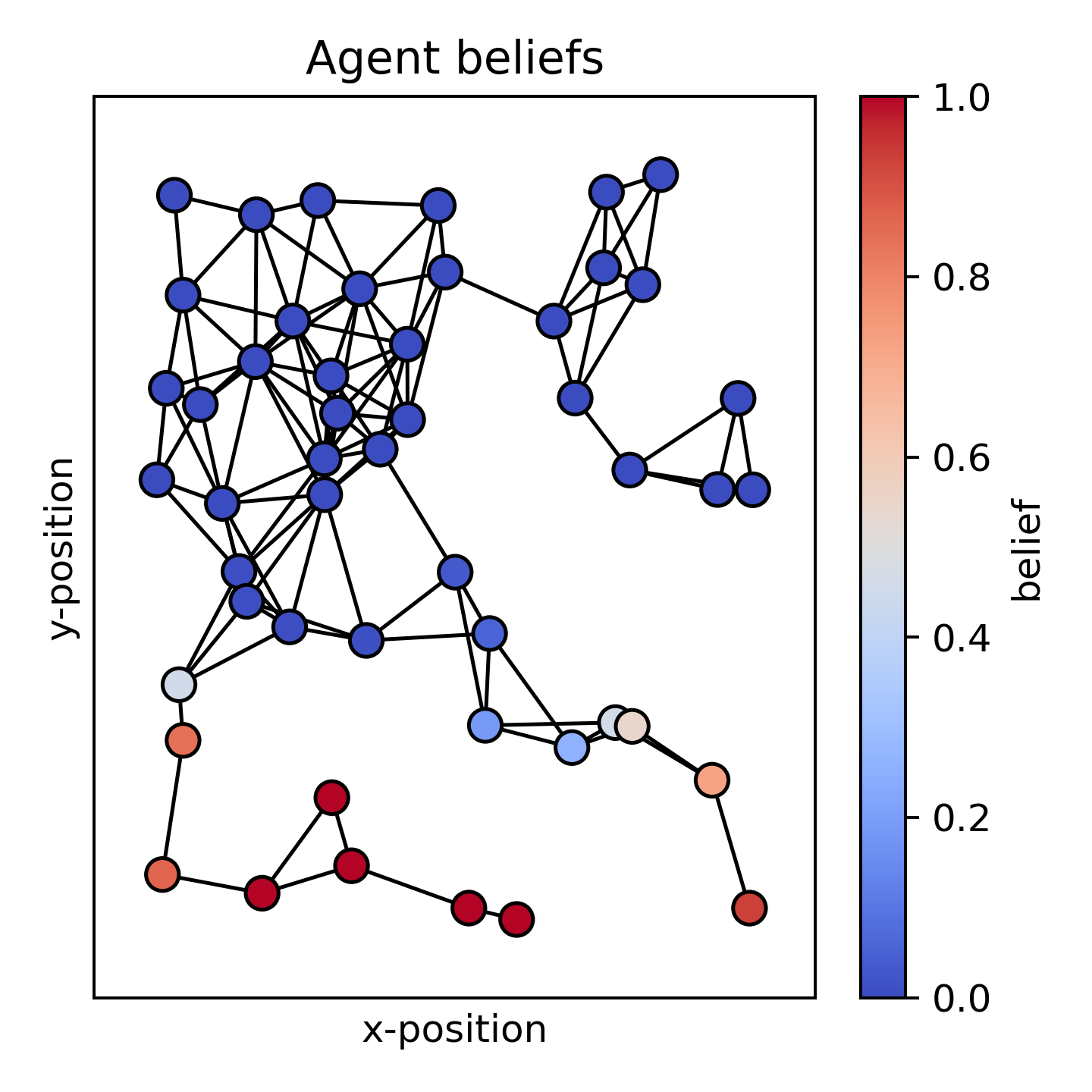}
        \caption{$r_g=0.2$}
    \end{subfigure}\\
    \begin{subfigure}{0.25\textwidth}
        \centering
         \includegraphics[trim={0 5mm 0 3mm},clip, width = \textwidth]{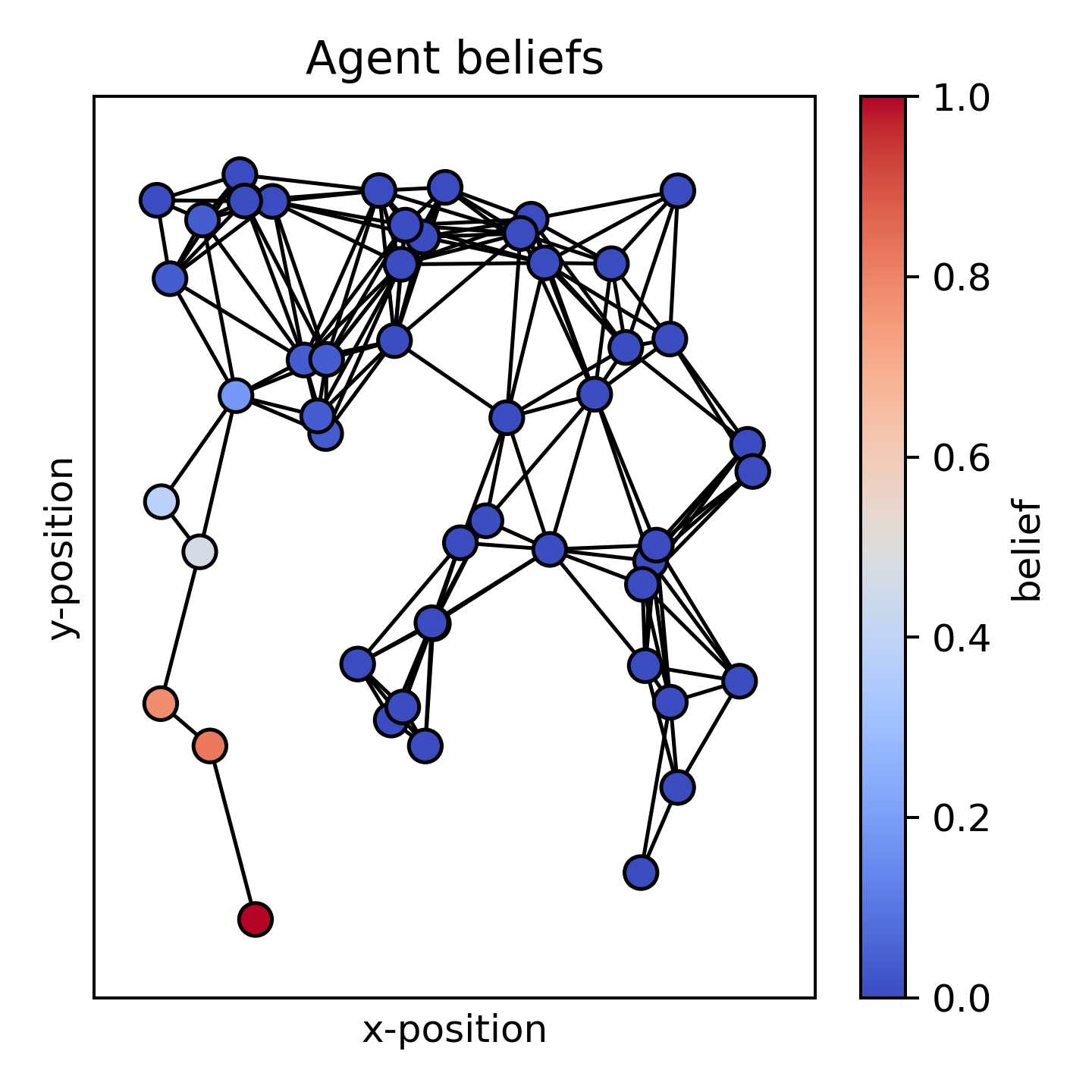}
         \caption{$r_g=0.25$}
    \end{subfigure}%
    \begin{subfigure}{0.25\textwidth}
        \centering
         \includegraphics[trim={0 5mm 0 3mm},clip, width = \textwidth]{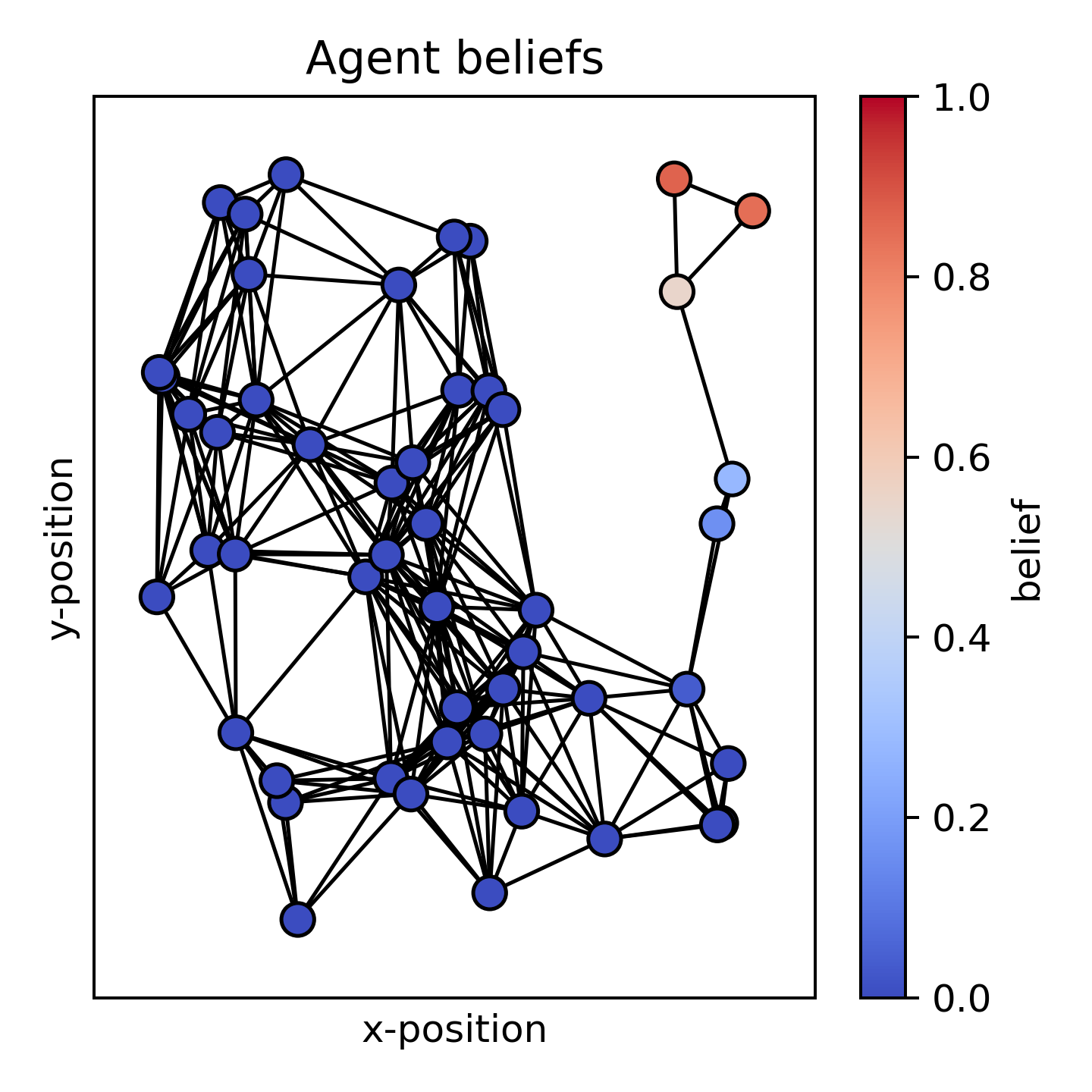}
         \caption{$r_g=0.3$}
    \end{subfigure}%

    \caption{State at termination of simulation for iterations in which convergence was not reached. This illustrates how the agent belief before absorption in a state of complete or zero trust appears localized, with high trust in regions with few connections, and low trust in regions with many connections.}
    \label{fig:noncon_states}
\end{figure}

We depict the network and the belief of agents for illustrative simulation runs that did not converge in simulated time in Figure~\ref{fig:noncon_states}. Networks that do not reach convergence in simulated time have regions which are dense, as well as sparse. In the dense regions, belief is low, and in the sparse regions belief is high.
In Figure~\ref{fig:noncon_states} we see how in the densely connected regions of the network trust is low, while in the sparsely connected regions trust is high. This illustrates that metastable disagreement is a phenomena that emerges with network structure\footnote{By the nature of $\lambda$ taking values in $\mathbb{R}$, the probability that $x^{k-1}<1/\lambda$ is increasing in $x\in[0,1]$ and decreasing in $k\in\{2,N\}$. Thus, regardless of distribution of $\lambda$, we expect a cutoff in the group-size under (above) which contribution (defection) is favored. This would result in similar patterns observed in Figures~\ref{fig:corr}, and~\ref{fig:noncon_states} simply shifted depending on where the cutoff is.}. We know that if we could observe the system for an infinite time, then eventually by chance the beliefs and actions of all agents will converge to either all contribute or all-defect state. In the non-equilibrium states, agent belief is highly dependent on the local network structure.

\subsection{Regular graphs, and metastability in geometric graphs}\label{sec:results_reg}
To disentangle the effect of the complex network topology and the effect of interaction size, we also study the dynamics on circulant graphs. We consider 50 agents spaced on a circle, and connect each agent to its $l$ nearest neighbors\footnote{These correspond to the circulants $C_{50}^{1},C_{50}^{1,2},C_{50}^{1,2,3}$ and $C_{50}^{1,2,3,4}$.} for $l\in\{2,4,6,8\}$. Each agent has the same degree, allowing us to study the effect of interaction size without heterogeneity in density. We keep the CDF $F(x)=1/x^4$, and $\alpha=0.3$. The proportion of simulation runs which converge to the always-contribute and the always-defect steady state is tabulated in Table~\ref{tab:numcon_reg}. In contrast to the simulation runs on the random geometric graphs, all the simulation runs converge within the simulated time.

\begin{table}[htb]
    \centering
    \begin{tabular}{c|c|c|c|c}
         $l$ &  $2$ & $4$ & $6$ & $8$\\
         \hline
         Contribution & 1 & 0.508 & 0 & 0 \\
         Defection & 0 &  0.492 & 1 & 1
    \end{tabular}
        \caption{Portion of runs out of 500 which converged to contribution and defection in $10^7$ rounds.}
        \vspace*{-2em}
    \label{tab:numcon_reg}
\end{table}
For the circulant connecting each agent to $l$ nearest neighbors, the games played by the agents are always of size $k=l+1$. Thus for $l=4$ the an agent with belief $x$ contributes with probability $x$. Conversely when $l<4$ ($l>4$) an agent with belief $x$ cooperates at probability $y>x$ ($y<x$) thus favoring contribution (defection) which is reflected in the results of the simulation tabulated in Table~\ref{tab:numcon_reg}.

\begin{figure}[htb]
    \flushleft
        \includegraphics[trim={0 5mm 0 3mm}, clip, width = 0.4\textwidth]{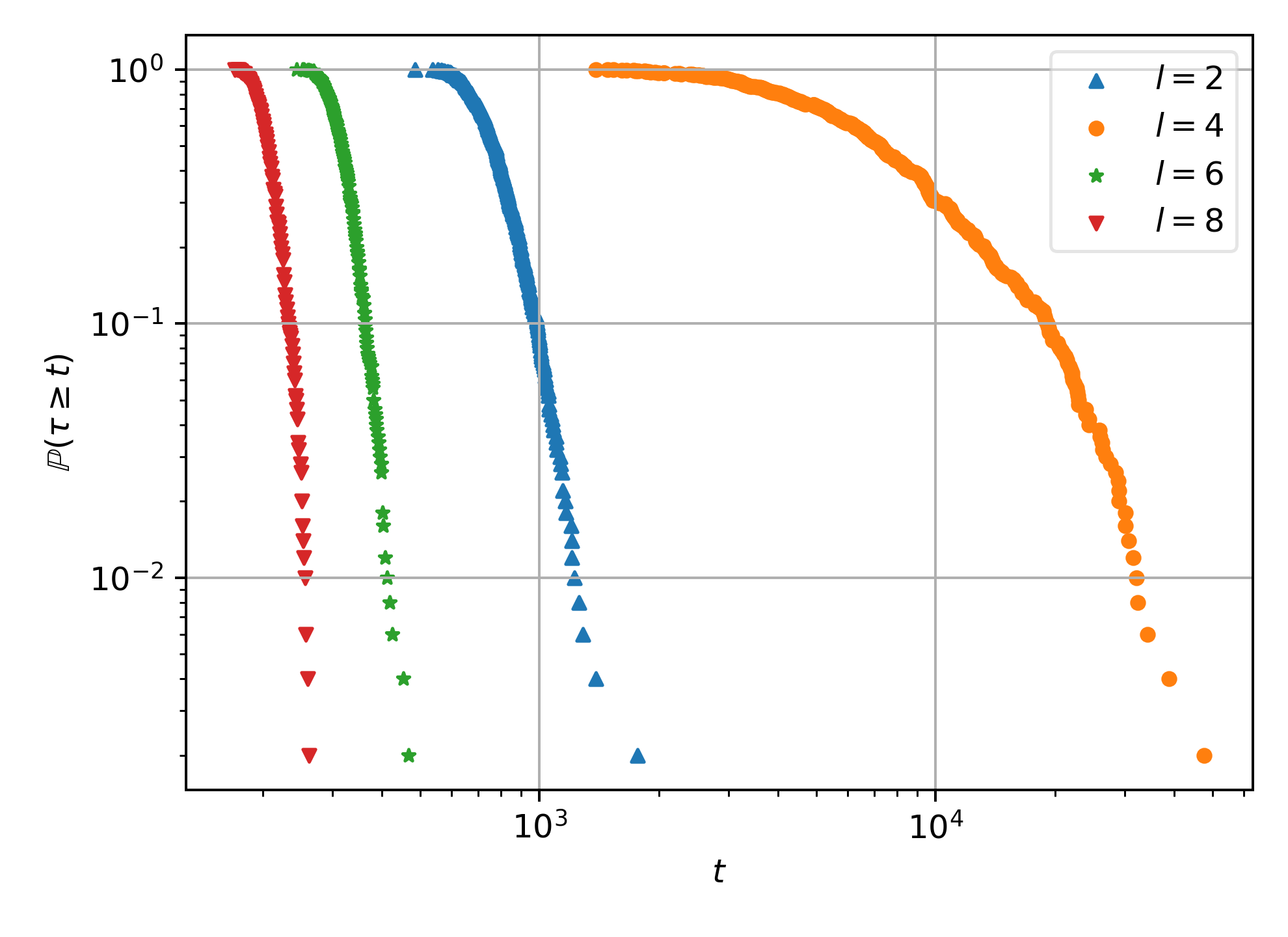}
    \caption{Tail probabilities ($\mathbb{P}(\tau\geq t)$) (on a log-log scale) for the time to convergence for different values of nearest neighbors $l$.}
    \label{fig:con_reg}
\end{figure}

The tail probabilities of the time to convergence $\tau$ for the regular graphs are plotted in Figure~\ref{fig:con_reg} on a log-log scale. In this case $\tau$ is not heavy-tailed. Furthermore, as $l$ increases from 2 to 4 the time to convergence increases, and then decreases again from 4 to 6 and 6 to 8. We expect that for $F(x)=1/x^m$ a similar pattern would emerge with an increase until $l=m$, and a subsequent decrease. The dynamics are fastest when the games are either smallest or largest. At the threshold $l=m$ they would be slowest to converge, because then neither steady-state is favored, as illustrated in Table~\ref{tab:numcon_reg}.

The difference in tail distributions seen when comparing Figure~\ref{fig:con} and Figure~\ref{fig:con_reg} indicates the behavior of the dynamics is fundamentally different in these two cases. We posit this difference is due to metastable states when the population structure takes the random geometric graph form. To test and illustrate this we run additional simulation runs paying attention to the total belief of the agents in the population. If a state of the agent beliefs is metastable we expect $\sum_{i=1}^N x_i(t)$ to reach some value and subsequently fluctuate (due to the randomness in the system) around this value before eventually being absorbed in one of the two states $\sum_{i=1}^N x_i(\tau)\approx 0$ or $\sum_{i=1}^N x_i(\tau)\approx N$. When running these simulations, we change the maximum running time to $10^6$ and vary $\alpha \in \{0.1,0.3\}$. We choose a shorter running time to avoid overly long simulation runs.
We vary the value of $\alpha$ to illustrate that the behavior is qualitatively the same.

\begin{figure}
    \centering
    \begin{subfigure}{0.45\textwidth}
        \centering
            \includegraphics[trim={0 5mm 0 4mm},clip, width = \textwidth]{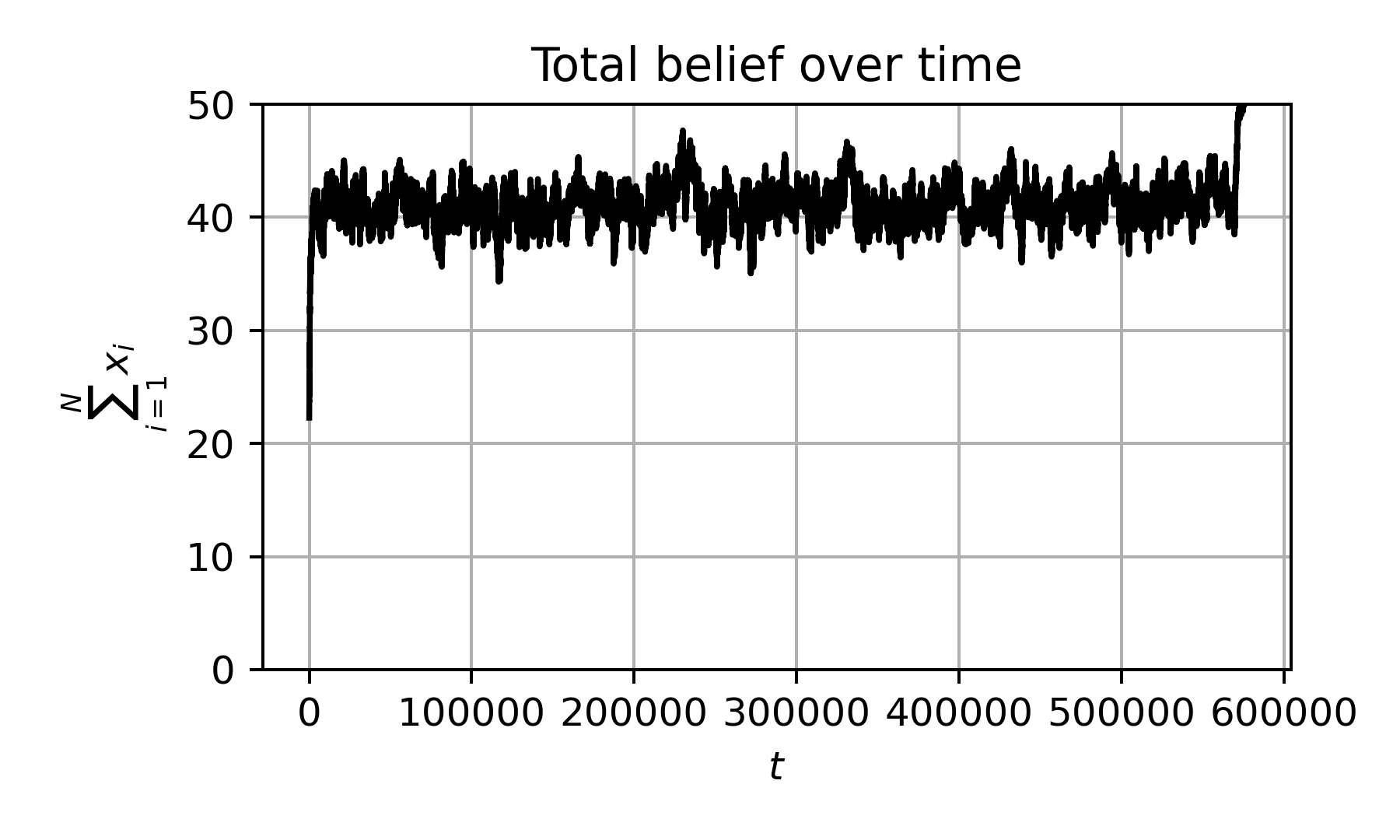}
            \caption{$\alpha=0.1$}\label{fig:a10}
    \end{subfigure}\\%
    \begin{subfigure}{0.45\textwidth}
        \centering
         \includegraphics[trim={0 5mm 0 4mm},clip,width = \textwidth]{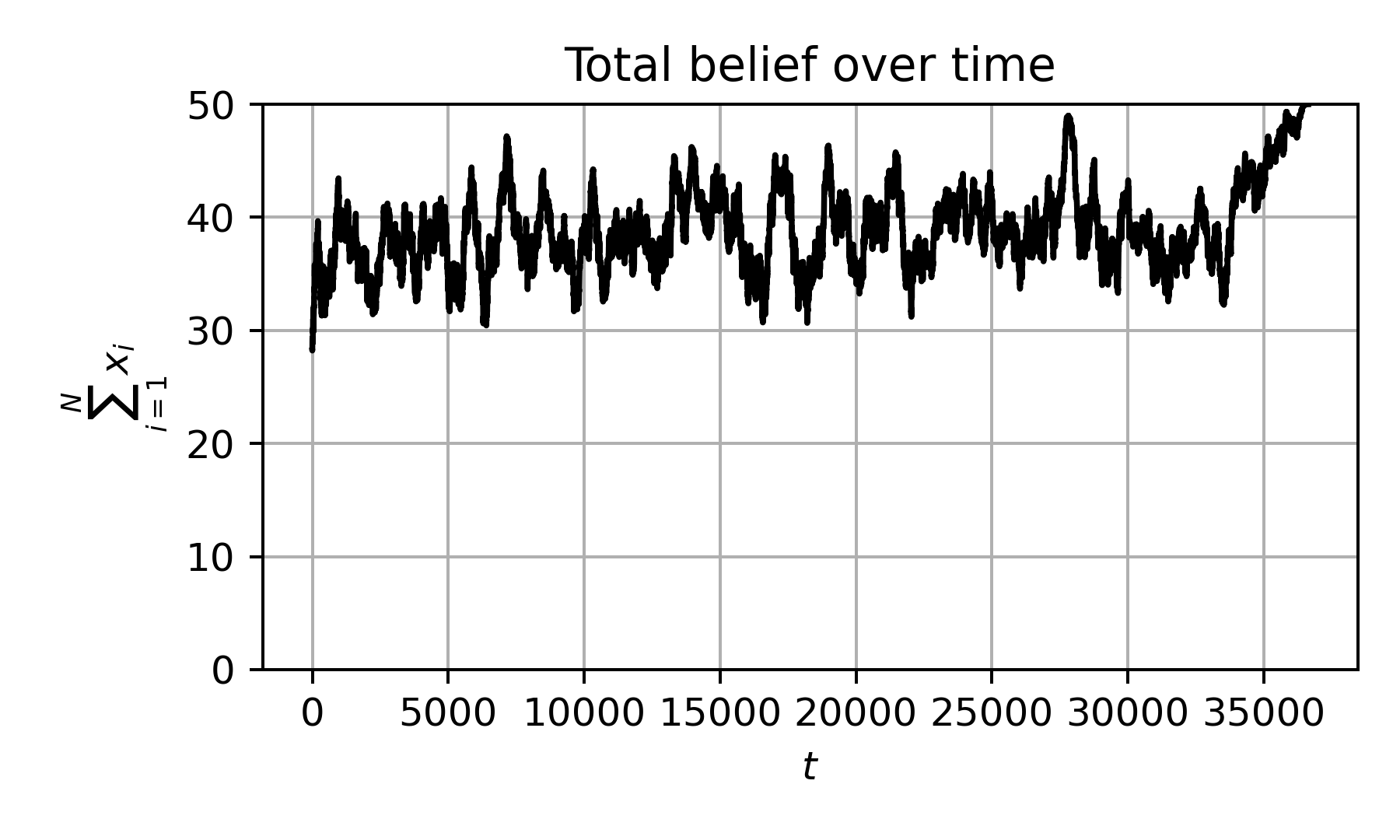}
         \caption{$\alpha=0.3$}\label{fig:a30}
    \end{subfigure}%
    \caption{The total agent belief over time for simulation runs in which we posit a metastable state is reached in finite time and eventually exited before $t=10^6$. In these simulations $N=50$, $r_g=0.15$ and $\alpha$ is indicated in the subcaption.}
    \label{fig:meta_belief}
\end{figure}

We plot the total population belief over time for two runs (one each for $\alpha=0.1$ and $\alpha = 0.3$) which illustrate metastable behavior in Figure~\ref{fig:meta_belief}. The total belief quickly reaches a value around which fluctuation occur for a finite time before absorbing in the steady state where $\bm{x}\approx 1^N$. Observe that the fluctuations are larger for bigger $\alpha$. This is sensible because $\alpha$ is the learning step size; how much the random fluctuations of the most recent round weigh in the agent belief. 

\begin{figure}
    \centering
    \begin{subfigure}{0.245\textwidth}
        \centering
            \includegraphics[trim={0 5mm 0 4mm},clip, width = \textwidth]{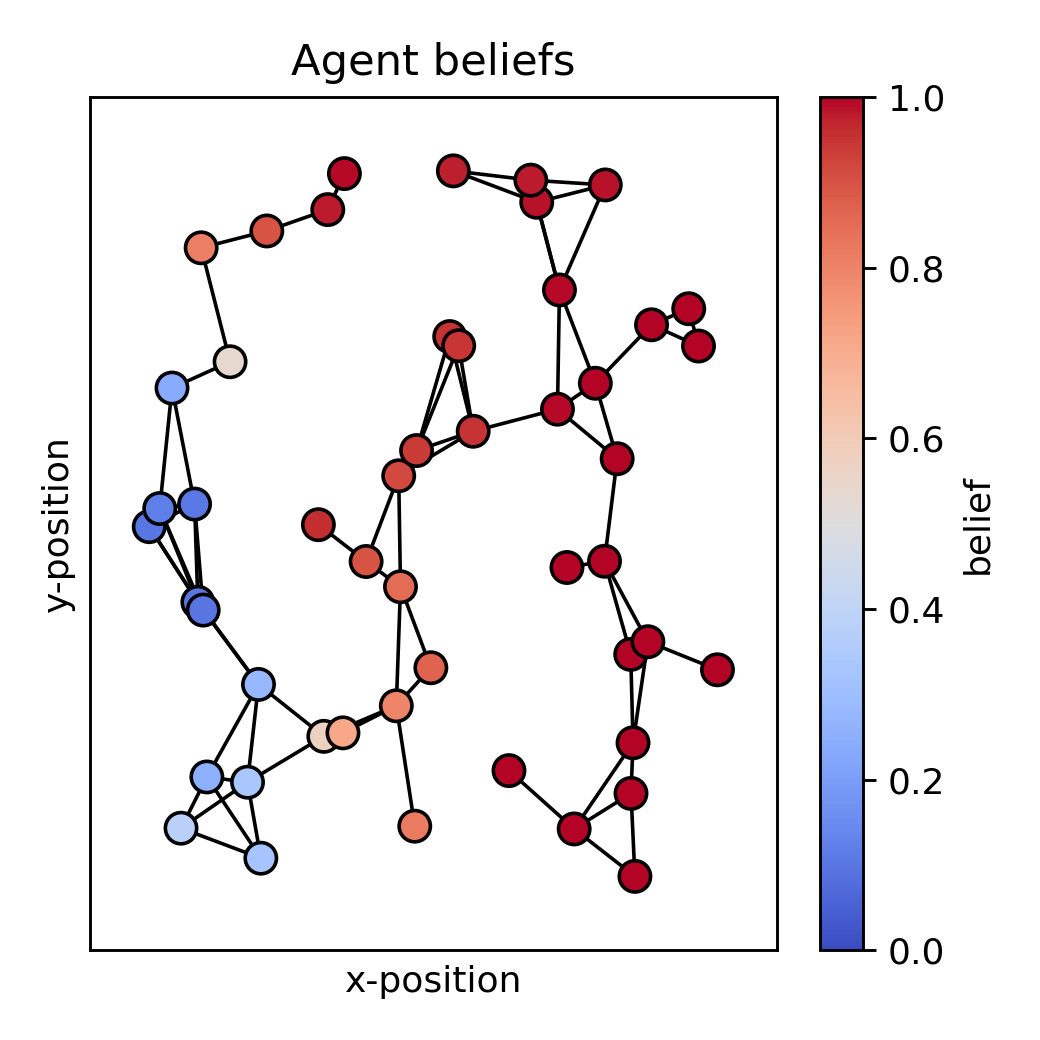}
            \caption{$\alpha=0.1$}\label{fig:a10_net}
    \end{subfigure}%
    \begin{subfigure}{0.245\textwidth}
        \centering
         \includegraphics[trim={0 5mm 0 4mm},clip, width = \textwidth]{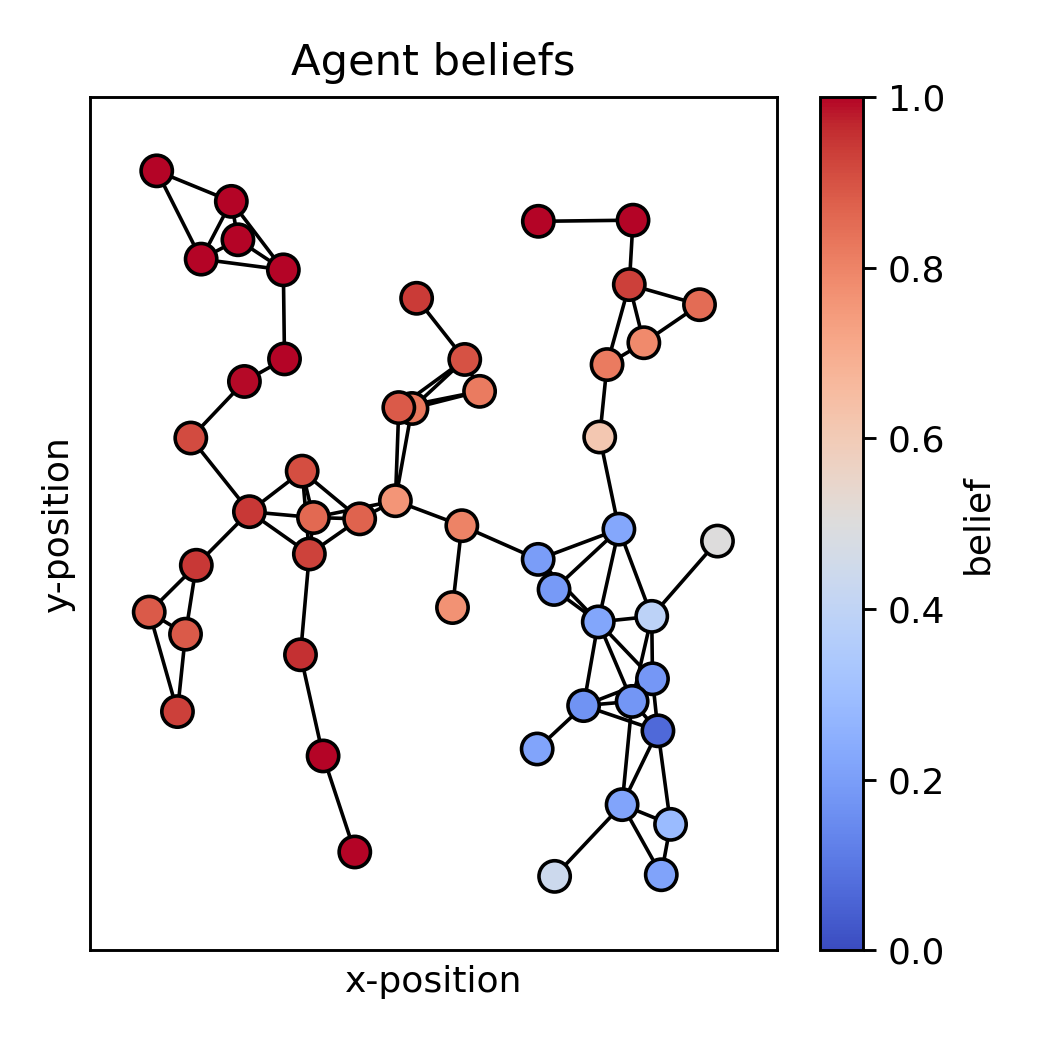}
         \caption{$\alpha=0.3$}\label{fig:a30_net}
    \end{subfigure}%
    \caption{Representative agent beliefs from the metastable period illustrated in Figure~\ref{fig:meta_belief}. In these simulations $N=50$, $r_g=0.15$ and $\alpha$ is indicated in the subcaption.}
    \label{fig:meta_net}
\end{figure}

We plot the network and agent beliefs of a representative round during the metastable period of these two illustrative runs in Figure~\ref{fig:meta_net}. These states are similar to the states plotted in Figure~\ref{fig:noncon_states} in having a region with low density (smaller game size and more trust) and a region with high density (larger game size and less trust). This is not possible in regular graphs and thus the most likely explanation for the metastability observed in random geometric networks.

\section{Conclusion}\label{sec:conc}
We close with a summary of our results and a discussion of directions for future work. 
\subsection{Summary and discussion}
In this paper we have defined a model for an all-or-nothing version of the public goods game with random payoffs to be played on networks. The agents in the model use the adaptive learning rule called exponential moving average to estimate the behavior of the agents in their neighborhood of the network. This learning rule is adaptive because the weight given to the most recent observation does not diminish with time. 

The main theoretical result of the paper is Proposition~\ref{prop:converge} which states that eventually a steady state must be reached. The steady state in question either involves all agents always playing the contribute action or all agents always playing the defect action. 

By simulation we study the effect of complex networks on the time and destination of this convergence. In particular we studied the effect of the radius in the random geometric graph model. We see that a larger radius (which implies more connections) results in more convergence to the defect steady state. In agreement with this, when simulating the dynamics on regular graphs, a larger interaction size results in more runs resulting in long-term defection. In comparison to the regular network, the complex network structure slows down convergence, as recently demonstrated for opinion dynamics~\cite{Zarei2024}. 

By the nature of the distribution of the time to convergence we posit that the system admits metastable states in which both contribution and defection take place. Iterations which do not converge in the simulated time have heterogeneous topology including regions of high and low density. This is confirmed when we inspect the total (summed) agent belief which may reach metastable value (with fluctuations) for a long period of time before `jumping' to one of the truly stable states.


\subsection{Future work}
The theoretical result in this paper is only proved for the exponential moving average learning rule. It would be good contribution to classify learning rules that do and do not result in asymptotic convergence to steady state of a pure strategy on connected networks. We suspect that when the underlying game admits pure Nash equilibria, these may be learned by a variety of algorithms also in the network setting. {Conversely it is interesting to investigate whether changing the learning rate $\alpha$ to be decreasing with $t$ would stabilize the states posited to be metastable in this paper.}

We posit metastability of the non-converged dynamics in this model and have illustrated this with examples. With more computational power it may be possible to conduct experiments which elucidate the reasons behind the possibility of metastable non-convergence. The question arises: `What network characteristics (local or global) are required to allow metastable states with positive probability?' We suggest looking into things such heterogeneity of the local characteristics in the graph.


A massive body of literature exists on evolutionary dynamics on regular graphs such as grids, and tori. In this paper we observe very different behavior on geometric networks (intermediate disagreement) and regular networks (fast convergence to steady-state). This highlights how important it is to consider the effects of non-regular network topology on learning and evolutionary dynamics. {The networks considered in this paper are highly stylized, thus illustrating a proof of concept. It is left for future work to delve deeper into the consequences of more realistic network structures.}


\begin{acks}
This research was supported by the European Union’s Horizon 2020 research and innovation programme under the Marie Skłodowska-Curie grant agreement no. 945045, and by the NWO Gravitation project NETWORKS under grant no. 024.002.003. \includegraphics[height=1em]{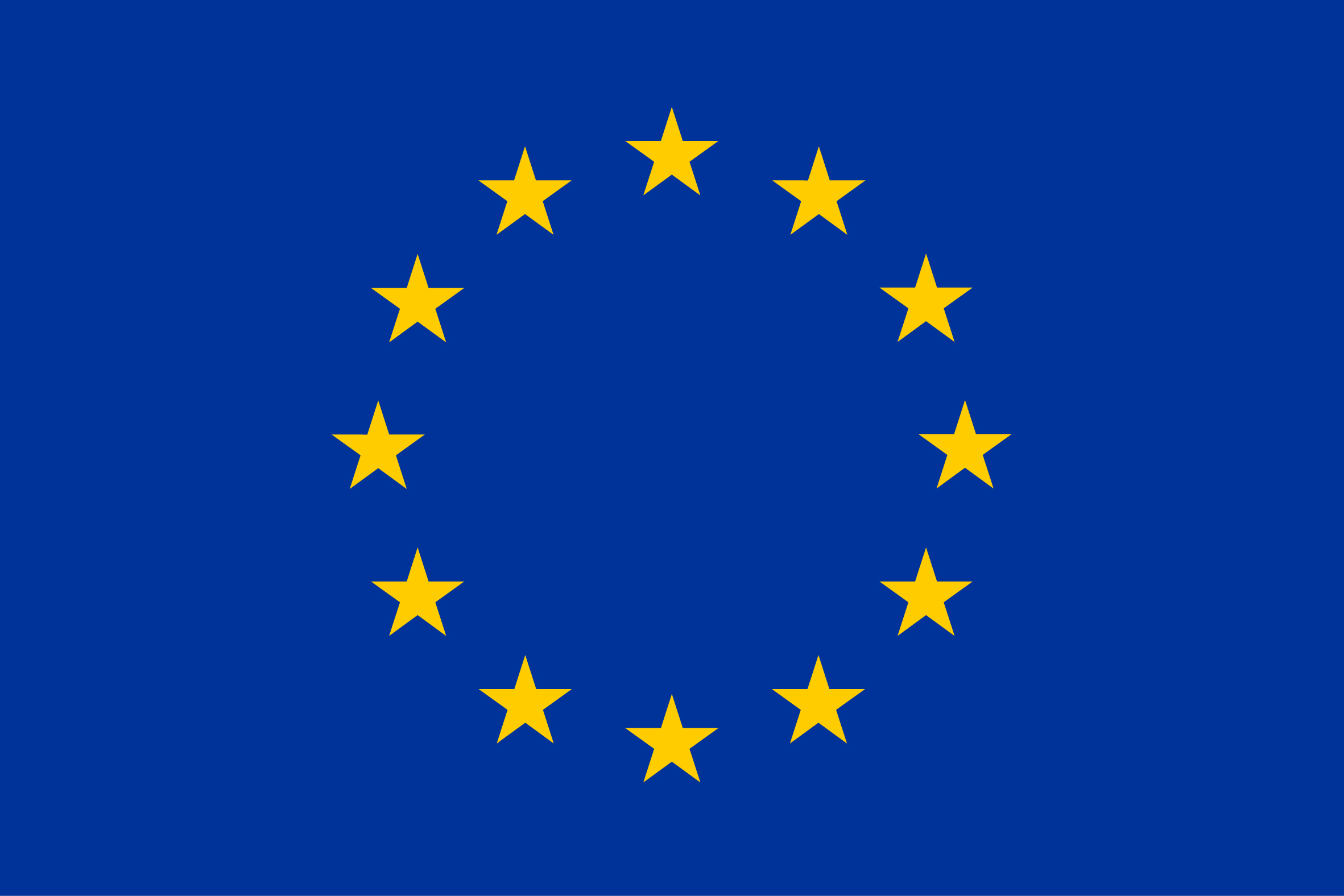}

The data for Figures 2, 3 and 5 are available at \href{https://github.com/Benephfer/All-or-nothing_Public_Goods}{Github: Benephfer/All-or-nothing\_Public\_Goods}. 
\end{acks}

\bibliographystyle{ieeetr} 
\bibliography{refs} 

\appendix

\section{Intuitive model explanation}
The technical model description is required for the statement of Proposition~\ref{prop:converge} and the presentation of the results of the simulation. Here we provide a more intuitive explanation of the model. 

\subsection{Agent belief and decision making}
The agents in the model are endowed with a belief $x$ which has the interpretation: `the probability at which they believe a random neighbor will contribute in a round of the public goods game.' When an agent is drawn to play the game, they are aware of the size of the interaction (they know how many players there are in the round), and they observe their private reward for that round $\lambda_i(t)$. The motivation is that the reward $\lambda_i(t)$ may be something which the agents arrive upon internally (their own utility evaluation of certain outcomes). 

When deciding which action to take, the agent acts as if they calculate the joint probability of all other players in the game contributing. Using that joint probability one can calculate the expected value of taking the contribute action (knowing also the reward they would receive if all agents contribute). The agent takes the action with the greater expected value as if calculated for only that round (thus the agents may be described as short-sighted). 

\subsection{Agent learning}
Once all agents have decided which action to take, the payoffs are awarded. Note the agents do not use the payoff to learn like in traditional reinforcement learning methodologies. The agents observe the precise number of contribute and defect actions taken by the players of the game, and so use this observation to update their belief based on exponential moving average. Their new belief is the weighted average of their previous belief and their current observation. The weights are $1-\alpha$ for their old belief, and $\alpha$ for the most recent observation. 

\begin{figure*}[htb]
\includegraphics[scale = 0.8]{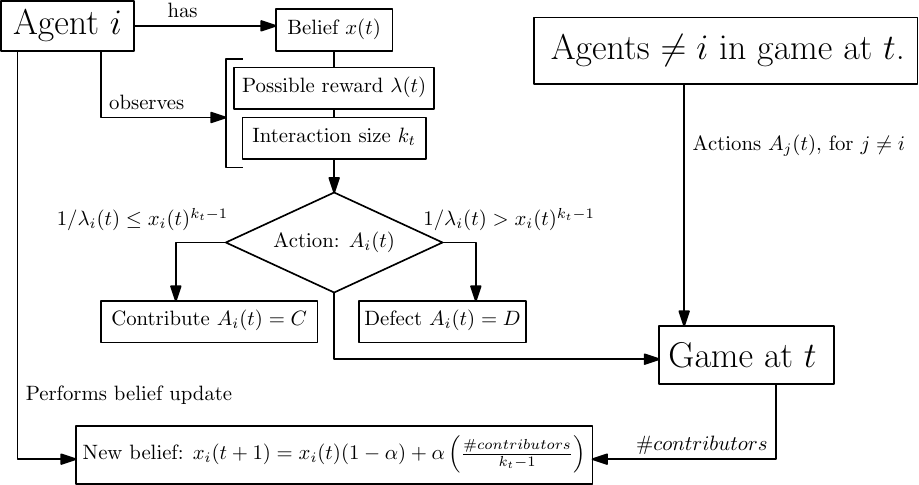}
\caption{The dynamics of the model illustrated for one round from the perspective of agent $i$. Note this is the same for all agents in the round, and thus described only from the perspective of one such agent. Note that the $\#contributors$ is calculated not including $i$ so that this matches (\ref{eq:update}).}
\label{fig:model}
\end{figure*}

This learning rule is dynamic because if one uses it to track a random variable which shifts over time, this shift is detected. This is made possible by keeping the weight of the most recent observation constant. The dynamics of the model are illustrated from the perspective of one representative agent ($i$) in Figure~\ref{fig:model}.

\subsection{Differences to traditional Reinforcement learning}

The learning rule used by the agents is not model-free. We assume the agents know the rules of the interaction as well as the size of the interaction when it happens. Furthermore, the agents' belief variable $x$ has a clear interpretation in the context of this model: The probability they believe an opponent will take the contribute action. This is different to the traditional reinforcement learning paradigm which is usually model-free. Such agents learn only by taking actions and observing rewards. They are not aware of how the actions they take and the actions taken by other agents interface with the rewards they get except through experience.

If one wishes to model a case when agents have less information about the environment (interaction size, payoff structure, etc) then it would make more sense to use a model-free learning rule such as Q-learning for example. 

A good starting point would be stateless, epsilon-greedy Q-learning. That is each agent has a belief vector (length two) in which they track the `quality' of taking each of the two actions (contribute, and defect). When drawn to interact they would take the action to which they have assigned the greater Q-value (or explore by taking the alternative action at probability $\epsilon>0$).

\end{document}